\documentclass[a4paper,11pt]{article}
\usepackage{jheppub} 
\usepackage{lineno}
\usepackage{xspace}
\usepackage{xcolor}
\usepackage{comment}
\usepackage{subfigure}
\usepackage{mathrsfs}
\usepackage{csquotes}

\newcommand{\jpsi} {\ensuremath{{\mathrm J}/\psi}\xspace}

\newcommand{\pp}           {pp\xspace}

\newcommand{\PbPb}         {\mbox{Pb--Pb}\xspace}
\newcommand{\pA}           {\mbox{p--A}\xspace}
\newcommand{\AAcoll}           {\mbox{A--A}\xspace}
\newcommand{\pPb}          {\mbox{p--Pb}\xspace}

\newcommand{\pt}           {\ensuremath{p_{\rm T}}\xspace}

\newcommand{\nineH}        {$\sqrt{s}~=~0.9$~Te\kern-.1emV\xspace}
\newcommand{\seven}        {$\sqrt{s}~=~7$~Te\kern-.1emV\xspace}
\newcommand{\eight}        {$\sqrt{s}~=~8$~Te\kern-.1emV\xspace}
\newcommand{\twoH}         {$\sqrt{s}~=~0.2$~Te\kern-.1emV\xspace}
\newcommand{\twosevensix}  {$\sqrt{s}~=~2.76$~Te\kern-.1emV\xspace}
\newcommand{\five}         {$\sqrt{s}~=~5.02$~Te\kern-.1emV\xspace}
\newcommand{\fiveRunThree}         {$\sqrt{s_{\mathrm{NN}}}~=~5.36$~Te\kern-.1emV\xspace}
\newcommand{\fiveExactly}  {$\sqrt{s}~=~5$~Te\kern-.1emV\xspace}
\newcommand{\twosevensixnn}{$\sqrt{s_{\mathrm{NN}}}~=~2.76$~Te\kern-.1emV\xspace}
\newcommand{\fivenn}       {$\sqrt{s_{\mathrm{NN}}}~=~5.02$~Te\kern-.1emV\xspace}

\newcommand{\GeVc}         {Ge\kern-.1emV/$c$\xspace}
\newcommand{\MeVc}         {Me\kern-.1emV/$c$\xspace}
\newcommand{\TeV}          {Te\kern-.1emV\xspace}
\newcommand{\GeV}          {Ge\kern-.1emV\xspace}
\newcommand{\GeVtwo}       {Ge\kern-.1emV$^2$\xspace}
\newcommand{\MeV}          {Me\kern-.1emV\xspace}
\newcommand{\GeVmass}      {Ge\kern-.1emV/$c^2$\xspace}
\newcommand{\MeVmass}      {Me\kern-.1emV/$c^2$\xspace}

\title{\boldmath Femtoscopy-driven searches for saturated gluonic matter in inclusive photonuclear processes}

\author[1]{S. Ragoni}
\author[2]{, P. Chakraborty}
\author[2]{, A. Kisiel}
\author[2]{, G. Kornakov}
\author[1]{, S. Pulawski}
\affiliation[1]{University of Silesia in Katowice,\\ 75 Pułku Piechoty 1A, \\ 42-500 Chorzów, Poland, PL}
\affiliation[2]{Faculty of Physics, Warsaw University of Technology,\\ Koszykowa 75, \\ 00-662 Warszawa, Poland, PL}

\emailAdd{simone.ragoni@cern.ch}
\emailAdd{pritam.chakraborty@pw.edu.pl}
\emailAdd{adam.kisiel@pw.edu.pl}
\emailAdd{georgy.kornakov@pw.edu.pl}
\emailAdd{szymon.pulawski@us.edu.pl}

\abstract{We present femtoscopy as a new way to search for saturated gluonic matter in inclusive photonuclear processes, such as inclusive ultraperipheral collisions at the Large Hadron Collider (LHC) and the inclusive photonuclear reactions at the Electron-Ion Collider (EIC). As the femtoscopic approaches are sensitive to the space-time structure of the particle emitting source, they are ideal in providing insights also about the initial stage of the collision, where the gluon distributions may impact the effective size of the dipole the quasireal photons oscillate into. This technique demonstrates its capabilities in isolating nuclear shadowing and gluon saturation effects. Finally, we show how a femtoscopic approach is highly sensitive to sub-fermi scale structures typically observed in ultraperipheral collisions, such as gluonic hot spots. }

\begin{document}
\maketitle
\flushbottom

\section{Introduction}
Photonuclear processes have been studied extensively at several colliders, ranging from HERA, to more recently the Large Hadron Collider (LHC) and the Relativistic Heavy Ion Collider (RHIC), and the next generation measurements will also be possible using the data that will be available at the Electron-Ion Collider (EIC). Within the context of heavy-ion collisions, ultraperipheral collisions (UPCs) rose as a promising venue for studying gluon distributions, through the measurement of exclusive vector meson production processes. Both RHIC and LHC experiments have published a substantial amount of differential studies regarding exclusive photoproduction and lepton pair production \cite{alice-vector-meson, cms-vector-meson, lhcb-vector-meson, ATLAS:2022ryk, star-vector-meson}. In particular, UPCs are used to study both nuclear shadowing and gluon saturation phenomena \cite{ALICE:2020ugp, STAR:2007elq, ALICE:2023jgu, Acharya:2748581, LHCb:2022ahs}. The ALICE Collaboration has reported the first observation of dense saturated gluonic matter \cite{ALICE:2023gcs}, where the $t$-dependence of incoherent \jpsi production favours models including subnucleonic degrees of freedom.

More recently, inelastic photonuclear interactions have shown great promise to extend the studies in \pp, \pA and \AAcoll by using the photon to probe the nucleus instead. This results in a complementary field of study to that typically explored with exclusive photonuclear measurements. ATLAS~\cite{ATLAS:2021jhn, ATLAS:2025tof} and ALICE~\cite{Ragoni:2024dlh} results demonstrate how inelastic photonuclear events open a new venue to study collective effects in smaller collision systems, bridging the gap between hadronic, heavy-ion collisions and electron--nucleus collisions at the EIC. 

While hadronic collisions have been studied extensively using several techniques, which also highlight the initial geometry of the event, inclusive photonuclear processes have mostly been studied in terms of \pt-distributions mostly. For instance, femtoscopy has been used successfully as a way to study the source size and particle production mechanisms typically involved in heavy-ion collisions~\cite{Goldhaber:1960sf,Kopylov:1972qw,Kopylov:1973qq,Cocconi:1974pr,Kopylov:1974th,Gyulassy:1979yi,Lednicky:1979ig,Pratt:1986cc,Makhlin:1987gm,Hama:1987xv,Sinyukov:1989xz,Sinyukov:1994vg,Akkelin:1995gh,Lisa2005,Lednicky:2005af}, and to infer the characteristics of the source which arises in \pPb~\cite{PhysRevC.91.034906} and \pp collisions~\cite{PhysRevD.84.112004}. Measurements at the LHC and RHIC have enabled systematic femtoscopic studies of source sizes, emission dynamics, and hadron–hadron interactions across different collision systems from \pp to hadronic \pA and \AAcoll. The connection of these findings to $e^+e^-$ collisions~\cite{Achard2011} is yet to be established. 

In this paper, we will show how femtoscopic techniques applied to inelastic photonuclear events, which will also be referred to below as inelastic or inclusive UPCs, succeed in giving insights on the nucleonic structure of the nucleus. Femtoscopy reveals itself as a viable technique to investigate the size of subnucleonic fluctuations, and thus address the core predictions of saturation theories. The results shown here may be applied directly to ultraperipheral collisions and to the photonuclear data that will become available with the EIC.


\section{Femtoscopy within the photonuclear context}


While femtoscopic techniques can readily be applied to any context to build correlation functions, their interpretations require clear considerations regarding the case under examination. The two particle momentum-correlation function can be defined as the ratio of probability ($P_{1,2}$) of simultaneous detection of two particles having momentum $p_1$ and $p_2$, respectively, to the product of the single-particle detection probabilities ($P_{1}$ and $P_{2}$) as given in Eq.~\ref{eq:CF_basic}~\cite{PhysRevC.92.054908,PhysRevC.93.024905}:
\begin{eqnarray}
	C(p_{\rm 1}, p_{\rm 2}) = \frac{P_{1,2}(p_{\rm 1}, p_{\rm 2})}{P_1(p_{\rm 1})P_2(p_{\rm 2})}. 
    \label{eq:CF_basic}
\end{eqnarray}
In experiments, the correlation function of two particles is constructed by taking the ratio of the distribution of correlated pairs produced in same events (\enquote{signal}) to the uncorrelated-pair distribution stemming from different events (\enquote{background}) as given in Eq.~\ref{eq:CF_expt}~\cite{PhysRevC.92.054908}:
\begin{eqnarray}
 C(\mathbf{q}^*)=T(\mathbf{q}^*)/M({\mathbf{q}^*}) \text{,}
 \label{eq:CF_expt}
\end{eqnarray}
where, $\mathbf{q}=\mathbf{p}_1 - \mathbf{p}_2$, and the asterisk (*) corresponds to the pair-rest frame  of reference (PRF), where the total momentum of pair is zero. The reconstructed correlation function can be related to the underlying dynamics of the particle-emitting source function using the Koonin–Pratt equation as given in Eq.~\ref{eq:koonin}~\cite{PhysRevC.92.054908}
\begin{eqnarray}    
    C(q^*) = \int d^3r^* \  S(r^*) |\Psi(r^*, q^*)|^2 \label{eq:koonin}
\end{eqnarray}
where, $q^*$=$|\mathbf{q^*}|$, $r^*$ is the relative pair separation. The $\Psi(r^*, q^*)$ is referred to as the pair-wave function and corresponds to the interactions between the particles in a pair with relative separation $r^*$ and relative momentum $q^*$. For a pair of identical particles, it includes quantum-statistical (QS) (anti)symmetrization and final-state interaction (FSI) i.e. Coulomb and/or strong interaction. The $S(r^*)$ is referred to as the source function and corresponds to the probability density of the directionally averaged pair-relative separation. However, within the context of photonuclear physics, the space-time structure of the source function needs to be generalised using the Bjorken-$x$ of the interaction, and thus the most general form of the source function is $S(r^*, x)$. In UPC physics the source that needs to be encoded in the Koonin-Pratt equation has to take into account the quasireal photon flux, the fluctuation of the photon in a diquark, and the interaction of the diquark with the gluon distribution of the target. An expression of the unnormalised source $\tilde{S}(r^*, x)$ is:
\begin{equation}
\tilde{S}(r^*, x)
\;\propto\;
\Biggl[
\underbrace{\int_{b>2R_A}\! d^2 b \int d\omega\; N_\gamma(\omega,b)}_{\text{EPA photon flux}}
\;\times\;
\underbrace{S_{\text{shad}}(x,b)}_{\text{shadowing}}
\Biggr]
\;
\underbrace{\int_0^1 dz\; \bigl|\Psi_{\gamma\rightarrow q\bar q}(r^*,z)\bigr|^2}_{\text{dipole}}
\;
\underbrace{\Bigl[\,1 - e^{-\,r^{*2} Q_s^2(x)/4}\Bigr]}_{\text{gluon saturation}} \text{ ,}
\label{eq:source}
\end{equation}
where $\int_{b>2R_A}\! d^2 b \int d\omega\; N_\gamma(\omega,b)$ incorporates the photon flux with the UPC condition that the impact parameter is larger than the sum of the nuclear radii from the EPA approach, ${S_{\text{shad}}(x,b)}$ is the nuclear shadowing factor, $\int_0^1 dz\; \bigl|\Psi_{\gamma\rightarrow q\bar q}(r^*,z)\bigr|^2$ encodes the fluctuation of the photon to a diquark, and finally we use $1 - e^{-\,r^{*2} Q_s^2(x)/4}$ as the parametrization for the gluon saturation regime. Each nucleus acts as a source of quasi-real photons whose impact parameter $b$–dependent flux is modeled starting from the point‐charge EPA flux:
\begin{equation}
  n_{\rm pt}(\omega,b)
  = \frac{Z^2\alpha}{\pi^2\,b^2}\,\xi^2
    \Bigl[K_1(\xi)^2 + \tfrac{1}{\gamma^2}\,K_0(\xi)^2\Bigr],
  \quad
  \xi \equiv \frac{\omega\,b}{\gamma}\,,
  \label{eq:point_flux}
\end{equation}
where $Z$ is the nuclear charge, $\alpha$ the fine‐structure constant, $\gamma$ the Lorentz factor, and $K_{0,1}$ the modified Bessel functions,
and the nuclear shadowing and gluon saturation terms are also separated.

For a spherically symmetric source with size $R_{inv}$, the source function can be modeled as a one-dimensional Gaussian function as given in Eq.~\ref{eq:source_1D}~\cite{PhysRevC.92.054908}: 
\begin{eqnarray}
    S(r^*)\sim exp
    \left(-\frac{{r^*}^2}{4R^2_{inv}}\right) \label{eq:source_1D},
\end{eqnarray}
and the two-particle correlation function can be approximated analytically for using the Bowler-Sinyukov formula as given in Eq.~\ref{eq:bowler}~\cite{BOWLER199169,SINYUKOV1998248,PhysRevC.93.024905,PhysRevC.92.054908}:
\begin{eqnarray}    
    C(q) = N[(1-\lambda)+\lambda\cdot K\left(q, R_{inv}\right)\cdot\left(1+\exp(-R_{inv}^2\cdot q^2)\right)],
    \label{eq:bowler}
 \end{eqnarray}
where $N$ is the normalisation constant, the $\lambda$ corresponds to the strength of correlation , $K\left(q, R_{inv}\right)$ refers to the Coulomb effect calculated at $q$ for $R_{inv}$. Under this approximation, any possible $x$-dependence of the source function should be observed in the extracted values of $R_{inv}$.



\section{Analysis of Monte Carlo data}
We have applied the formalism shown above to Monte Carlo simulations generated using STARlight~\cite{Klein:2016yzr} interfaced with DPMJET III 19.3.7~\cite{Djuvsland:2010qs}, which will be henceforth referred to as STARlight+DPMJET. The events are categorised into two classes based on the charged-particle multiplicity measured in the forward pseudorapidity region ($|\eta|>$3.0): 1--200 and 200--400. Although UPC events in the experiments are expected to have very low multiplicities, this study has the liberty to carry out the calculations in relatively higher values of multiplicity ranges due to dealing with Monte Carlo data at the generated level.

In each multiplicity class, the identically charged pions are selected from the mid-pseudorapidity region ($|\eta|<$0.8) and are paired with each other to construct the correlation function. To calculate the kinematics and the relative separation, the pair is first boosted from the laboratory frame to the Longitudinally Co-Moving System (LCMS), defined by three directions: out (parallel to the pair transverse momentum), long (along the beam direction), and side (orthogonal to the other two). In LCMS, the total momentum of the pair along the long direction vanishes. The pair is finally boosted to PRF, in which the pair-wave function is computed. Since STARlight+DPMJET does not contain any interaction between the particles in a pair at the generated level, it is added to each pair of the signal distribution via the \enquote{afterburner} technique. The pair-wave function for identical charged pions is symmetrised according to the Bose–Einstein statistics shown in Eq.~\ref{eq:Psi_QS}~\cite{PhysRevC.90.064914}:
\begin{eqnarray}
    \Psi_{\mathrm{QS}} = 1 + cos(2k^*r^*) \label{eq:Psi_QS}
\end{eqnarray}
where $\mathbf{k}^*=\mathbf{q}^*/2$, and also includes final state interactions (FSI) arising from both the Coulomb repulsion shown in Eq.~\ref{eq:Psi_FSI}~\cite{tlzz-sks4}, and strong interaction. However, the effect of strong interaction is negligible for identical pion pairs and hence is not considered in this study.
\begin{eqnarray}
	\Psi_{\mathrm{FSI}}=\sqrt{A_C(\eta_{\mathrm{FSI}})}\left[e^{-ik^*r^*}F(-i\eta_{\mathrm{FSI}},1,i\zeta)\right],  \label{eq:Psi_FSI}
\end{eqnarray}
where $A_C$ is the Gamow penetration factor, $\eta_{\mathrm{FSI}}=1/(k^*a_c)$, $a_c=387.25$ fm is the Bohr radius of the pion-pion pair, $F$ is the confluent hypergeometric function, $\zeta=k^*r^*(1+\cos\theta^*)$, and $\theta^*$ is the angle between $\mathbf{k}^*$ and $\mathbf{r}^*$. The product of $\Psi_{\mathrm{QS}}^2$ and $\Psi_{\mathrm{FSI}}^2$ calculated using Eq.~\ref{eq:Psi_QS} and Eq.~\ref{eq:Psi_FSI}, respectively, is added as a weight to each pair of the signal distribution. 

\begin{figure}[!h]
\centering
\includegraphics[width=1.\columnwidth]{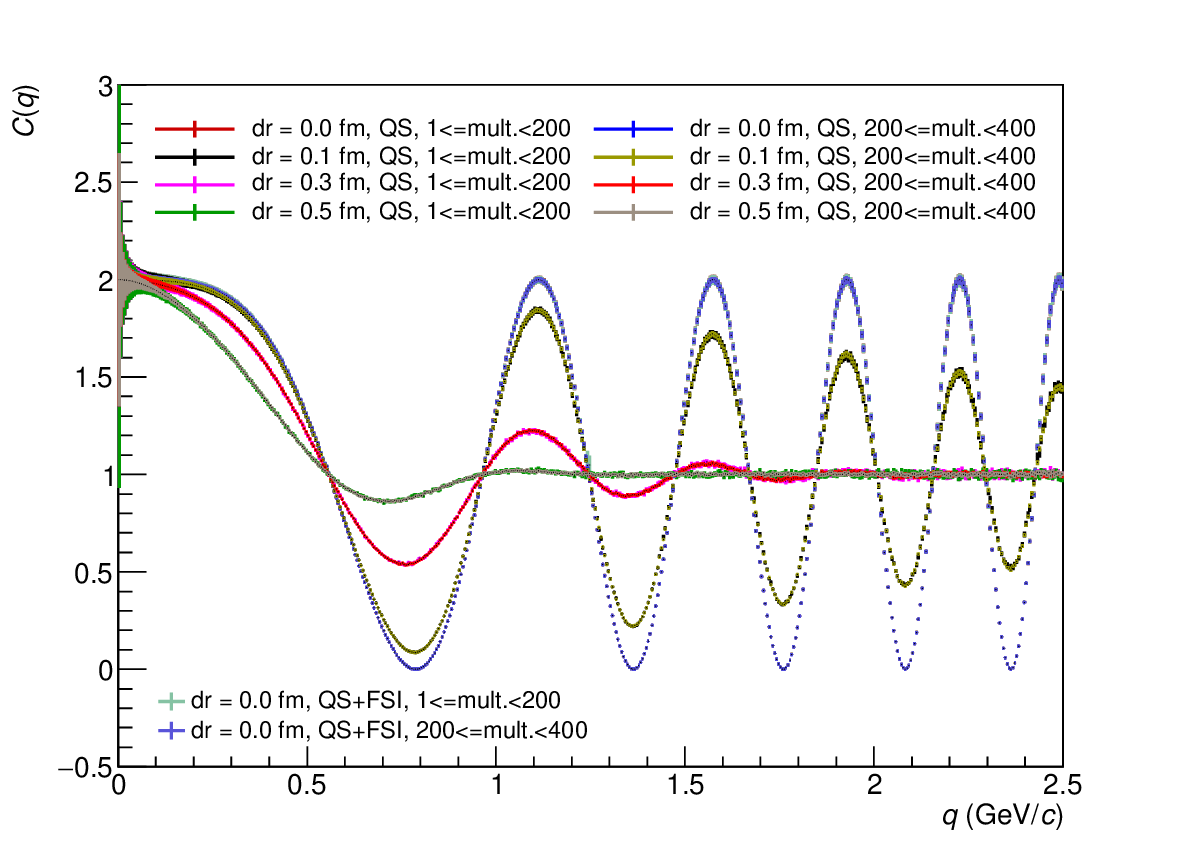}
\caption{\label{fig:femto-mc} (Color online) Two-particle correlation functions of identical pions generated by introducing a set of additional displacements ($dr$) in the separation in a pair, analysed in two forward-multiplicity classes of Pb--Pb UPC events at \fiveRunThree generated with STARlight+DPMJET.} 
\end{figure}

Moreover, to account for an additional broadening of the distributions of relative-pair separation, a Gaussian smearing is applied independently to each LCMS component of the relative separation vector. The out, side, and long components are modified by adding random displacements sampled from one-dimensional normal distributions with fixed standard deviations as shown in Eq.~\ref{eq:add_r}: 
\begin{eqnarray}
    r_i' = r_i + dr_i \label{eq:add_r}
\end{eqnarray}
where, $i=$ out, side, and long. The modified components ($r_i'$) of pair-separation vector is subsequently boosted to the PRF to calculate $r^*$ and the corresponding pair weights. The identical-pion correlation functions produced in inclusive UPC events using STARlight+DPMJET and calculated using different values of additional displacement, $dr=0.0$, 0.1, 0.3, and 0.5 fm to the relative-pair separation are shown in Fig.~\ref{fig:femto-mc}. For any given value of $dr$, the correlation functions for different values of multiplicity are compatible with each other within uncertainties. For $dr=0.0$, the correlation functions are calculated including both quantum statistics (QS) and QS+FSI contributions. It is observed that the effect of FSI on the shape of the correlation function is negligible, as the QS+FSI result nearly overlaps with the QS-only result. Therefore, for the remaining values of $dr$, only QS is considered. The correlation functions generated without any additional displacement demonstrate the feature of a typical point-like source. However, introducing an additional displacement in the pair separation essentially increases the effective source size of the emission region. Consequently, the momentum-interference pattern in the correlation function becomes suppressed and broadened at high and low $q$ region, respectively, and this feature becomes more prominent with increasing effective width of the source, as shown in Fig.~\ref{fig:femto-mc}, which is consistent with the expected femtoscopic sensitivity of the correlation function to the spatial size of the source. The observed suppression is most pronounced for $dr=0.5$ fm, where the correlation function shows a clear undershooting below unity at intermediate $q$. A qualitatively similar feature has been reported in measurements of Bose–Einstein correlations in hadronic Z-boson decays at LEP. This shows that femtoscopy is a viable technique to not only qualitatively but also quantitatively investigate the gluonic fluctuations and gluonic hot spots. Therefore, femtoscopy is already sensitive to sub-fermi level structures such as those that emerge from gluon saturation theory \cite{ALbacete:2010ad} and, more recently, from new findings from ultra-peripheral collisions at RHIC and the LHC \cite{ALICE:2023gcs, Mantysaari:2016ykx, Cepila:2025lbr}. These results demonstrate that a dedicated fit to the correlation functions obtainable with inclusive UPC and photonuclear data will be quintessential in understanding the origin of the onset of saturated regimes, and hold promise in being the simplest observable to resolve the sub-fermi effects that characterise this type of investigations. 


\section{Approximations in the UPC case}
Owing to the well-defined correlation functions shown in Fig.~\ref{fig:femto-mc}, it is also possible to extract the effective radius directly from the correlation function itself. By means of the one-dimensional Gaussian approximation, the effective radius $R_{eff}$ will be given by:
\begin{equation}
    R_{eff}^2(x) = -\frac{1}{\lambda}\cdot \frac{d^2}{dr^2} [C(q, x)-1] \text{.}
\end{equation}
We use the typical DIS formalism to the rapidity $y$ as a proxy for $x$, by means of:
\begin{equation}
    x\approx \frac{2m_{\rm T}}{\sqrt{s_{NN}}}e^{\pm y} \text{ ,}
    \label{eq:dis}
\end{equation}
where $m_{\rm T} = \sqrt{m^2+\pt^2}$ is the transverse mass, $\sqrt{s_{NN}}$ is the centre-of-mass energy of the collision system, and the factor $2$ takes into account the usage of a pair of identical particles. While femtoscopy is typically applied to midrapidity acceptances, the ALICE detector allows for the additional possibility of measuring identical pion pairs even at forward rapidities, by means of the Muon Forward Tracker (MFT). Using the pion mass as input to Eq.~\ref{eq:dis}, this means that ALICE is potentially sensitive to $x_{mid}\sim 10^{-4}$ and $x_{MFT}\sim 10^{-6}$ for midrapidity and MFT rapidity-wise acceptances, respectively.

In the weak-saturation regime~\cite{Golec-Biernat:1998zce, Iancu:2003ge}, it is possible to expand the saturation term in Eq.~\ref{eq:source}. Assuming spherical symmetry, we work in a one-dimensional representation of the source similar to Eq.~\ref{eq:source}, parametrised as:
\begin{equation}
\tilde{S}_0(r)\propto \exp\!\left(-\frac{r^2}{4R_0^2}\right) \text{ ,}
\end{equation}
where $R_0$ sets the characteristic transverse size of the emission region, so that the unnormalised source in Eq.~\ref{eq:source} can be expressed in the weak-saturation limit as:
\begin{equation}
\tilde{S}_{weak}(r)\propto r^2 \tilde{S}_0(r) \text{ .}
\end{equation}
The mean squared separation is defined as
\begin{equation}
\langle r^2\rangle_{\text{weak}}
=
\frac{\int d^3r\, r^2\, \tilde S_{\text{weak}}(r)}
     {\int d^3r\, \tilde S_{\text{weak}}(r)} \text{ .}
\label{eq:r2_weak_def}
\end{equation}
From which we can derive that the effective femtoscopic radius is given by:
\begin{equation}
R_{\text{weak}}
=
\sqrt{\frac{5}{6}} R_0 \text{ .}
\label{eq:R_weak}
\end{equation}
Thus the femtoscopic radius is roughly independent from the Bjorken-$x$ (within higher-order corrections in the weak-saturation expansion). Sensitivity is recovered once p--Pb and \PbPb UPCs are compared, and a ratio performed. In this case, the \pPb UPCs only feature saturation; thus the previous equation holds. In \PbPb , shadowing appears as well. Sensitivity can then be restored if nuclear effects induce an $r$-dependent deformation of the source. This is precisely what is expected from nuclear shadowing in the dipole picture: shadowing is not merely an overall suppression of yields, but preferentially suppresses large dipoles, hence modifying the shape of $S(r)$. This shows that the combined analysis of inclusive UPCs with both protons and lead nuclei targets, is useful to also isolate nuclear shadowing effects from different nuclear phenomena.

\begin{figure}[!h]
\centering
\includegraphics[width=0.9\textwidth]{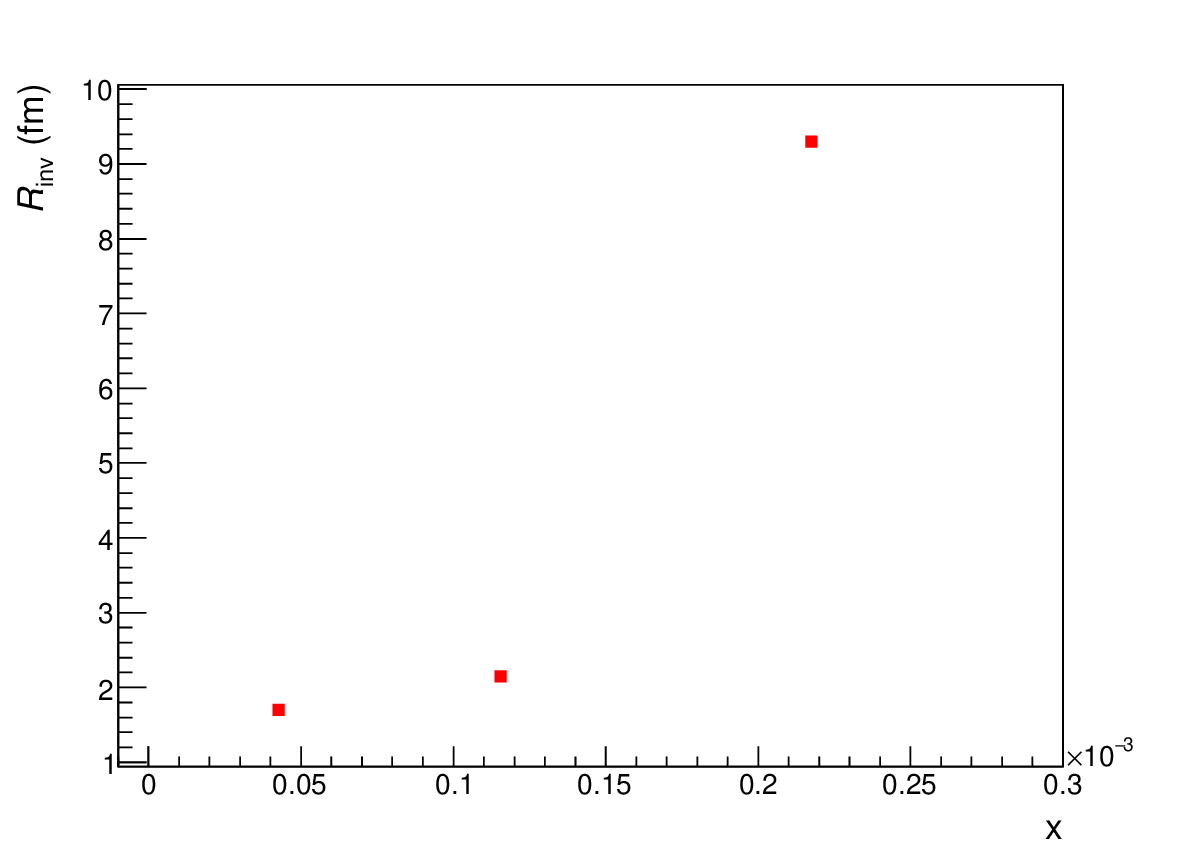}
\caption{\label{fig:R_x} (Color online) Effective width ($R_{\mathrm{inv}}$) of 1D source vs. Bjorken-x approximated from pp, p--Pb, and Pb--Pb collisions with ALICE.} 
\end{figure}

The measured width ($R_{\mathrm{inv}}$) of one dimensional sources measured using pion-pion femtoscopic correlations in pp, p--Pb, and Pb--Pb collisions at $\sqrt{s}=13$~TeV, $\sqrt{s_{\mathrm{NN}}}=5.02$~TeV, and $\sqrt{s_{\mathrm{NN}}}=2.76$~TeV, respectively, with ALICE are shown as the function of Bjorken-x, approximated using $m_{\mathrm{T}}=0.29$~GeV$/c$, in Fig.~\ref{fig:R_x}.

\section{Summary}
Femtoscopic techniques have been used extensively to investigate the source size that arises in hadronic heavy-ion collisions. In this paper we have shown how the technique can be applied to inclusive photonuclear processes, specifically speaking to inclusive ultra-peripheral collisions, which can already be measured at the LHC. Femtoscopic techniques are then able to give access to the sub-fermi structures that arise within gluon saturation theory. It is shown to hold promise to validate the predictions from gluon saturation theories, which predict sizes of the order of 0.1 to 0.4 fm, effects that result in a dampening of the correlation function itself. By doing a combined analysis of \pPb and \PbPb UPC data, for example, it is also possible to show that femtoscopy is able to directly isolate nuclear effects that originate from different phenomena. Finally, we have also shown how scaling of the source is dependent on both multiplicity and Bjorken-$x$, and shown how inclusive UPC data naturally extend the discussion of the results achieved in hadronic heavy-ion and pp collisions at colliders with the earlier results in ee collisions. The results shown here anticipate those that will become available by carrying out measurements with  EIC data, which will be used to extend the observations introduced herein.

\acknowledgments

This work was supported by the funds granted under the Research Excellence Initiative of the University of Silesia in Katowice, by the Research University – Excellence Initiative of Warsaw University of Technology via the strategic funds of the Priority Research Centre of High Energy Physics and Experimental Techniques, and by the Polish National Science Centre under agreement no. 2022/45/B/ST2/02029. 


\bibliographystyle{JHEP}
\bibliography{biblio.bib}






\end{document}